\documentclass[twocolumn,showpacs,superscriptaddress,eqsecnum,aps,prb]{revtex4-1}
\usepackage{amssymb}
\usepackage{graphicx}
\usepackage{bm}
\usepackage{epsfig}
\usepackage{epstopdf}
\usepackage{amsmath}
\usepackage{amstext}
\usepackage{float}

\begin{document}

\title{Cross-over from retro to specular Andreev reflections in bilayer
graphene}
\author{Dmitri K. Efetov}
\affiliation{Department of Electrical Engineering and Computer Science, Massachusetts
Institute of Technology, Cambridge, USA}
\author{Konstantin B. Efetov}
\affiliation{Theoretische Physik III, Ruhr-Universit\"{a}t Bochum, D-44780 Bochum, Germany}
\affiliation{National University of Science and Technology ``MISiS'', Moscow, 119049,
Russia}
\date{\today }

\begin{abstract}
Ongoing experimental progress in the preparation of ultra-clean
graphene/superconductor (SC) interfaces enabled the recent observation of
specular interband Andreev reflections (AR)\cite{andreev} at bilayer graphene (BLG)/NbSe$%
_{2}$ van der Waals interfaces [Nature Physics 12, (2016)\cite{dima}]. Motivated by
this experiment we theoretically study the differential conductance across a
BLG/SC interface at the continuous transition from high to ultra-low Fermi
energies $E_{F}$ in BLG. Using the Bogoliubov-deGennes equations\cite{bdg} and
the Blonder-Tinkham-Klapwijk formalism\cite{btk} we derive analytical expressions for
the differential conductance across the BLG/SC interface. We find a
characteristic signature of the cross-over from intra-band retro- (high $%
E_{F}$) to inter-band specular (low $E_{F}$) ARs, that manifests itself in a
strongly suppressed interfacial conductance when the excitation energy $%
|\varepsilon |=|E_{F}|<\Delta $ (the SC gap). The sharpness of these
conductance dips is strongly dependent on the size of the potential step at
the BLG/SC interface $U_{0}$.
\end{abstract}

\maketitle

\section{Introduction}

Andreev reflections\cite{andreev} (AR) at normal metal (N) to superconductor
(SC) interfaces describe the non-trivial conversion of a normal dissipative
current into a dissipationless supercurrent. When an electron from N is
injected onto a SC with an excitation energy $|\varepsilon |<\Delta $ (the
SC gap), it is reflected back as a hole with an exactly opposite direction
of motion. This perfect retro-reflection process is understood as the result
of the quasi-elasticity and momentum conservation of the process, combined
with the fact that the hole has a negative mass. This picture is however
only an approximation\cite{beenakker}. Due to a small energy loss of the electron due to its
condensation into a cooper pair, the holes energy and in-plane momentum, and
hence its angle of reflection, are in all generality smaller than that of
the incident electron. This effect is exceedingly small in typical metallic
systems where $E_{F}$ strongly exceeds typical energies of $\Delta $, and
perfect retro-AR in these systems has been confirmed in great detail \cite%
{angle}.

Recent experimental efforts however shifted towards more exotic materials
such as semi-conductors\cite{majorana}, topological insulators\cite{topo}, quantum Hall systems\cite{qhe},
carbon-nanotubes\cite{JD} and graphene\cite{heersche,dima}. Unlike typical metals, these can have much
lower $E_{F}$ where the retro-reflection process can be dramatically
altered. Single (SLG) and bilayer graphene (BLG) \cite{geim,kim} with their
semi-metallic band-structure and the continuous gate tunability of $E_{F}$
are uniquely suited to study the $E_{F}$ dependence of ARs at the cross-over
from large to arbitrarily small $E_{F}$. In the recently achieved regime
where $|E_{F}|<\Delta $\cite{dima}, the reflected hole can undergo an
inter-band transition from the conduction into the valence band, causing the
holes mass to change its sign. Under these conditions, energy and momentum
conservation dictate that the reflection angle becomes a non-trivial
function of $E_{F}$, resulting in a specular AR process for $E_{F}=0$ where
the angle of incidence and reflection are equal $\alpha=\alpha^{\prime}$ (Fig.1 (a) and (b)).

\begin{figure}[tbp]
\centering
\includegraphics[width=0.5\textwidth]{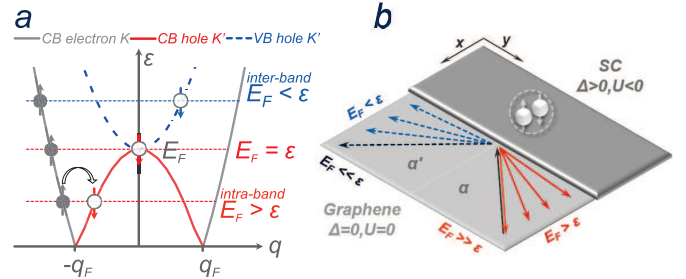}
\caption[Conductance for varying interfacial potential $U_{0}$]{(a)
Excitation spectrum $\protect\varepsilon(q)$ for a fixed $E_{F} < \Delta$.
With an increasing excitation voltage $\protect\varepsilon$, the momentum in
the y-direction $q$ of the reflected hole continuously increases from
negative to positive values, passing through zero when $E_{F}$=$\protect%
\varepsilon$. (b) Schematics of the reflection angles of AR holes in the
various energy limits. Starting from perfect intra-band retro-reflection in
the high $E_{F}$ limit, the reflection angle $\protect\alpha^{\prime}$
continuously increases towards $\protect\pi/2$ as $E_{F}$ is lowered. At the
cross-over point separating intra-band and inter-band ARs, $E_{F}=\protect%
\varepsilon$, $\protect\alpha^{\prime}$ exhibits a jump to $-\protect\pi/2$,
which eventually results in perfect inter-band specular reflection $\protect%
\alpha$=$\protect\alpha^{\prime}$ when $E_{F}=0$. }
\end{figure}

In this paper we work out a detailed analytical description of the
differential conductance across a BLG/SC interface at the continuous
transition from high $|E_{F}|\gg \Delta $ to low $|E_{F}|<\Delta $ Fermi
energies. While direct measurements of scattering angles are experimentally
challenging, measurements of a non-linear conductance $G_{NS}(\varepsilon )$
as a function of the excitation energy $\varepsilon $ are easily
experimentally approachable and can illuminate the underlying scattering
processes at the N/SC interface. We find two distinct regimes for $%
|\varepsilon |<|E_{F}|$ and $|\varepsilon |>|E_{F}|$ where intra-band
retro-reflections and inter-band specular-reflection are taking place,
respectively. We identify conductance dips in $G_{NS}(\varepsilon )$ at the
energy condition $|\varepsilon |=|E_{F}|$, where the hole is reflected onto
the charge neutrality point. This line in the energy phase space separates
the two distinct regimes, so marking the cross-over from retro intra-band to
specular inter-band AR.

\section{Model and method}

We model the BLG/SC junction considering a setup similar to the one
presented in the experimental device of Ref. \cite{dima}, and similar to the
one used in the theoretical model in Ref. \cite{beenakker} (Fig. 1 (a)). We assume an
impurity free BLG sheet in the half plane $x>0$ that has an ideal electrical
contact with a SC lead at $x<0.$ The SC is modeled by a BLG sheet with an
induced SC gap $\Delta \left( x\right) $ that appears due to the SC
proximity effect. For a more realistic BLG/SC interface we also assume that
the SC contact induces an additional potential $U\left( x\right) $ due to
work-function matching of the BLG and the SC, as it is typically observed
for all graphene/metal interfaces. Assuming that only singlet SC is induced
and that only electrons of different valleys form Cooper pairs in the BLG,
we write the Bogoliubov-deGennes (BdG) equations\cite{bdg} that describe the
electron motion in the system in the standard form

\begin{equation}
\left(
\begin{array}{cc}
\mathcal{H}\left( \mathbf{\hat{K},}x\right) -E_{F} & \Delta \left( x\right)
\\
\Delta ^{\ast }\left( x\right) & E_{F}-\mathcal{H}\left( \mathbf{\hat{K},}%
x\right)%
\end{array}%
\right) \left(
\begin{array}{c}
u \\
v%
\end{array}%
\right) =\varepsilon \left(
\begin{array}{c}
u \\
v%
\end{array}%
\right) ,  \label{0a1}
\end{equation}%
where

\begin{equation}
\mathcal{H}\left( \mathbf{\hat{K},}x\right) =\mathcal{H}_{0}\left( \mathbf{%
\hat{K}}\right) +U\left( x\right)  \label{0a1a}
\end{equation}%
and $\mathbf{\hat{K}}$ is the momentum operator.

The Hamiltonian $\mathcal{H}_{0}\left( \mathbf{K}\right) $ of the normal BLG
for one valley is written as a $4\times 4$ matrix \cite{wallace,falko,guinea}

\begin{equation}
\mathcal{H}_{0}(\mathbf{K})\mathcal{=}\hslash v\left(
\begin{array}{cccc}
0 & Ke^{i\alpha } & -t_{\perp } & 0 \\
Ke^{-i\alpha } & 0 & 0 & 0 \\
-t_{\perp } & 0 & 0 & Ke^{-i\alpha } \\
0 & 0 & Ke^{ia} & 0%
\end{array}%
\right)  \label{0a2}
\end{equation}

where $\alpha \left( \mathbf{K}\right) =\arctan \left( q/k\right)$ and $K=%
\sqrt{k^{2}+q^{2}}$ ($q$ is the y-component and $k$ is the $x$-component of $%
\mathbf{K}$).

At the BLG/SC interface (x=0) we assume a sharp potential step of $U(x)$ and
$\Delta(x)$ :

\begin{equation}
U\left( x\right) =\left\{
\begin{array}{cc}
-U_{0}, & x<0 \\
0, & x>0%
\end{array}%
\right. ,  \label{0a3}
\end{equation}

\begin{equation}
\Delta \left( x\right) =\left\{
\begin{array}{cc}
\Delta _{0}e^{i\phi }, & x<0 \\
0, & x>0%
\end{array}%
\right. .  \label{0a4}
\end{equation}

This potential profile is clearly an oversimplification, since in the
experiment these quantities can scale rather smoothly at the interface.
While our assumption may result in a certain overestimation of the
transmission amplitude at the interface, we do not think that the dependence
of the conductance on $E_{F}$ and $\varepsilon $ can essentially be
different for a smoothed potential step, and believe that this approximation
should not change the basic picture of the conversion from retro- to
specular ARs.

In accordance with the experimentally known parameters, we assume that the
potential $U_{0}$ is larger than the energies $E_{F},\Delta _{0}$ and $%
\varepsilon $ but the largest energy in BLG is the coupling energy $\hslash
vt_{\perp }$ between the layers. Therefore we use the inequalities

\begin{equation}
\hslash vt_{\perp }\gg U_{0}\gg E_{F},\Delta _{0},\varepsilon.  \label{0a5}
\end{equation}

Here we want to specifically point out that the inequalities (\ref{0a5}) are
different from those used in Ref. \cite{ludwig}, which leads to quite
different approximations and final results.

One can calculate the differential conductance $G_{NS}(\varepsilon )$ for $%
|\varepsilon |<\Delta _{0}$ and $T=0$ by considering the scattering of
particles that are moving from right to left. Following the formalism of
Ref. \cite{btk} one can solve the equations (\ref{0a1}-\ref{0a5}) separately
for $x>0$ and $x<0$, then match these solutions at $x=0$, and so find the
normal reflection $r(\varepsilon ,\alpha )$ and Andreev reflection $%
r_{A}(\varepsilon ,\alpha )$ amplitudes for all incidence angles $\alpha $

\begin{equation}
\frac{G_{NS}(\varepsilon )}{G_{NN}(\varepsilon )}=\int_{0}^{\pi
/2}(1-r^{2}(\varepsilon ,\alpha )+r_{A}^{2}(\varepsilon ,\alpha ))\cos
\alpha d\alpha ,  \label{0a6a}
\end{equation}%
where $G_{NN}(\varepsilon )$ is the differential conductance across the
interface of two normal BLG sheets.

Such a calculation is not very difficult for SLG \cite{beenakker}, however a
corresponding calculation for BLG is considerably more cumbersome and it is
not easy to obtain explicit formulas in all regions of the variables $%
\varepsilon $ and $E_{F}$. Here we follow Beenakkers approach \cite{beenrev}
that allows one to express the conductance $G_{NS}(\varepsilon )$ in terms
of transmission $t(\varepsilon )$ and reflection $r(\varepsilon )$
amplitudes of the scattering on the interface between two normal metals, by
simply putting $\Delta _{0}=0$. We write the differential conductance $%
G_{NS}(\varepsilon )$ in the form :

\begin{equation}
G_{NS}(\varepsilon )=\frac{4e^{2}}{h}Tr[m(\varepsilon )m^{+}(\varepsilon )],
\label{0a7}
\end{equation}%
with

\begin{equation}
m(\varepsilon )=t_{12}(\varepsilon )[1-e^{-2i\beta }r_{22}^{\ast
}(-\varepsilon )r_{22}(\varepsilon )]^{-1}t_{21}^{\ast }(-\varepsilon ),
\label{0a8}
\end{equation}%
where $\beta =\arccos (\varepsilon /\Delta _{0})$.

The conductance $G_{NN}(\varepsilon ,E_{F})$ across two normal regions $1$
and $2$ reads

\begin{equation}
G_{NN}(\varepsilon ,E_{F})=\frac{4e^{2}}{h}Tr\left\vert t_{12}(\varepsilon
,E_{F})\right\vert ^{2},  \label{0a10}
\end{equation}%
where regions $1$ and $2$ correspond to N $(x>0)$ and SC $(x<0)$,
respectively. In other words, Eq. (\ref{0a7}-\ref{0a10}) show that
calculating the transmission $t_{12}(\varepsilon )$, $t_{21}(\varepsilon ),$
and reflection $r_{12}(\varepsilon ),$ $r_{21}(\varepsilon )$ amplitudes
(see Appendix) for right and left moving particles at the interface one
obtains the differential conductance $G_{NS}(\varepsilon )$ of the BLG/SC
interface. The trace $Tr$ over the scattering channels in Eqs. (\ref{0a7}, %
\ref{0a10}) reduces to the integration over the momentum $q$ parallel to the
interface. Eqs. (\ref{0a7}-\ref{0a10}) also demonstrate that, in order to
calculate the differential conductance $G_{NS}$, it is enough to understand
the scattering on the interface between two normal metals, which is a
considerably simpler task than calculating the differential conductance
using the original formula (\ref{0a6a}).

Generally, one can see from Eqs. (\ref{0a7}, \ref{0a8}) an important
difference between the conductance of SLG and that of BLG. For SLG the
energy spectrum follows the Dirac equations and hence the transmission
amplitude for an electron is of order one for any $U_{0}$ due to the so
called Klein effect. In contrary, the spectrum of BLG is approximately
quadratic and one can conclude by using standard formulas \cite{landau} for
scattering on a step function that the transmission amplitude $%
t_{12}(\varepsilon )$ decays proportionally to $U_{0}^{-1/2}$ at large $%
U_{0} $ and therefore is very small. For this reason, while the conductance $%
G_{NN} $ in BLG is proportional to $t^{2},$ the conductance $G_{NS}$ is
proportional to $t^{4},$ and hence can be much smaller than $G_{NN}.$ The
presence of the SC gap can therefore strongly reduce the conductance across
the interface.

In the next Section we present the main analytical formulas for the
conductance leaving details of the derivation for the Appendix.

\section{Analytical expressions for the differential conductance}

We have to specifically distinguish between two
distinct regimes :\\

A) $|\varepsilon| < |E_{F}|$ - the reflected hole has positive energy
(conduction band) and negative mass $\Rightarrow$ intra-band
retro-reflection.\newline

B) $|\varepsilon| > |E_{F}|$ - the reflected hole has negative energy
(valence band) and positive mass $\Rightarrow$ inter-band
specular-reflection.

\subsection{$G_{NS}\left(\protect\varepsilon \right) $ for $\protect%
|\varepsilon| <|E_{F}|$ (retro-reflection)}

Eqs. (\ref{0a7}-\ref{0a10}) can be rewritten using the integration over the
longitudinal component $q$ of the momentum instead of the trace over the
transversal channels. However, it is even more convenient to integrate over
the incident angle $\alpha $. The angle $\alpha $ corresponds to the energy $%
\varepsilon ,$ while the reflection angle $\alpha ^{\prime }$ corresponds to the
energy $-\varepsilon $. These angles are related to each other by the
condition that $q$ is conserved during the reflection process and therefore

\begin{equation}
\frac{\sin \alpha ^{\prime }}{\sin \alpha }=-\sqrt{\frac{E_{F}+\varepsilon }{%
E_{F}-\varepsilon }}.  \label{0a43}
\end{equation}

From this relation one can conclude that Andreev reflections are possible
for the angles $\left\vert \alpha \right\vert <\alpha _{c},$ where

\begin{equation}
\alpha _{c}=\arcsin \sqrt{\frac{E_{F}-\varepsilon }{E_{F}+\varepsilon }}.
\label{0a44}
\end{equation}

Writing

\begin{equation}
\Phi \left( \alpha \right) =\arcsin \left( \sin ^{2}\alpha \right) ,
\label{0a45}
\end{equation}

\begin{equation}
\Phi \left( \alpha ^{\prime }\right) =\arcsin \left( \sin ^{2}\alpha
^{\prime }\right) =\arcsin \left[ \left( \frac{E_{F}+\varepsilon }{%
E_{F}-\varepsilon }\right) \sin ^{2}\alpha \right]  \label{0a45a}
\end{equation}

the conductance $G_{NS}$ can be reduced to the form

\begin{equation}
G_{NS}\left( \varepsilon \right) =2G_{0}K_{0}\left( \varepsilon \right)
\int_{0}^{\alpha _{c}}\frac{Y_{\varepsilon }\left( \alpha ,\alpha ^{\prime
}\right) \cos \alpha }{2\left\vert X_{\varepsilon }\left( \alpha ,\alpha
^{\prime }\right) \right\vert ^{2}}d\alpha ,  \label{0a46}
\end{equation}

where

\begin{eqnarray}
&&Y_{\varepsilon }\left( \alpha ,\alpha ^{\prime }\right) =\left\vert
t_{21}\left( \varepsilon \right) \right\vert ^{2}\left\vert t_{21}\left(
-\varepsilon \right) \right\vert ^{2}  \label{0a47} \\
&=&16L\left( \varepsilon \right) L\left( -\varepsilon \right) \cos \alpha
\cos \alpha ^{\prime }\left( 1+\sin ^{2}\alpha \right) \left( 1+\sin
^{2}\alpha ^{\prime }\right) ,  \notag
\end{eqnarray}

\begin{equation}
L\left( \varepsilon \right) =\sqrt{\frac{\left\vert \varepsilon
+E_{F}\right\vert }{U_{0}}}\ll 1,  \label{e1}
\end{equation}%
and

\begin{equation}
K_{0}\left( \varepsilon \right) =\sqrt{\left\vert \varepsilon
+E_{F}\right\vert t_{\perp }/\hslash v}.  \label{e2}
\end{equation}

The function $X_{\varepsilon }\left( \alpha ,\alpha ^{\prime }\right) $
entering Eq. (\ref{0a46}) equals

\begin{equation}
X_{\varepsilon }\left( \alpha ,\alpha ^{\prime }\right) =\frac{1}{2}\left[
1-e^{-2i\beta }r_{22}^{\ast }\left( -\varepsilon \right) r_{22}\left(
\varepsilon \right) \right] ,  \label{0a48}
\end{equation}%
where $\beta $ is given by the expression
\begin{equation}
\beta =\arccos \left( \varepsilon /\Delta _{0}\right) .  \label{0a9}
\end{equation}

Approximating this function by lowest orders in $L\left( \varepsilon \right)
$ we write

\begin{eqnarray}
&&\left\vert X_{\varepsilon }\left( \alpha ,\alpha ^{\prime }\right)
\right\vert ^{2}=\sin ^{2}\beta -2\sin \beta \Big[L\left( \varepsilon
\right) \sqrt{1+\sin ^{2}\alpha }  \notag \\
&&\times \sin \left( \beta +\Phi \left( \alpha \right) \right) +L\left(
-\varepsilon \right) \sqrt{1+\sin ^{2}\alpha ^{\prime }}\sin \left( \beta
-\Phi \left( \alpha ^{\prime }\right) \right) \Big]  \notag \\
&&+L^{2}\left( \varepsilon \right) \left( 1+\sin ^{2}\alpha \right)
+L^{2}\left( -\varepsilon \right) \left( 1+\sin ^{2}\alpha ^{\prime }\right)
\notag \\
&&+2L\left( \varepsilon \right) L\left( -\varepsilon \right) \sqrt{\left(
1+\sin ^{2}\alpha \right) \left( 1+\sin ^{2}\alpha ^{\prime }\right) }
\notag \\
&&\times \Big[\cos \left( 2\beta -\Phi \left( \alpha ^{\prime }\right) +\Phi
\left( \alpha \right) \right) -2\sin \Phi \left( \alpha \right) \sin \Phi
\left( \alpha ^{\prime }\right) \Big]  \notag \\
&&  \label{0a50}
\end{eqnarray}

The quadratic terms in $L\left( \varepsilon \right) ,$ $L\left( -\varepsilon
\right) $ have been kept in $\left\vert X_{\varepsilon }\left( \alpha
,\alpha ^{\prime }\right) \right\vert ^{2}$ because they are the only
contributions that do not vanish at $\beta \rightarrow 0$. In principle,
there are also quadratic terms proportional to $\sin \beta $, but they can
be neglected.

\subsection{$G_{NS}\left( \protect\varepsilon \right) $ for $\protect%
|\varepsilon| > |E_{F}|$ (specular reflection)}

In this case the angles $\alpha $ and $\alpha ^{\prime }$ are related to
each other as

\begin{equation}
\frac{\sin \alpha ^{\prime }}{\sin \alpha }=\sqrt{\frac{\varepsilon +E_{F}}{%
\varepsilon -E_{F}}}  \label{0a51}
\end{equation}

and the critical angle $\alpha _{c}$ equals to

\begin{equation}
\alpha _{c}=\arcsin \sqrt{\frac{\varepsilon -E_{F}}{\varepsilon +E_{F}}}
\label{0a52}
\end{equation}

For $\Phi \left( \alpha \right) $ we have the same relation as in Eq. (\ref%
{0a45}) and obtain for $\Phi \left( \alpha ^{\prime }\right) $

\begin{equation}
\Phi \left( \alpha ^{\prime }\right) =\arcsin \left[ \left( \frac{%
\varepsilon +E_{F}}{\varepsilon -E_{F}}\right) \sin ^{2}\alpha \right]
\label{0a53}
\end{equation}%
\newline

The conductance $G_{NS}\left( \varepsilon \right) $ is then determined as
before by Eq. (\ref{0a46}). For the function $Y_{\varepsilon }\left( \alpha
,\alpha ^{\prime }\right) $ we therefore have

\begin{eqnarray}
&&Y_{\varepsilon }\left( \alpha ,\alpha ^{\prime }\right) =\left\vert
t_{21}\left( \varepsilon \right) \right\vert ^{2}\left\vert t_{21}\left(
-\varepsilon \right) \right\vert ^{2}  \label{0a54} \\
&=&16L\left( \varepsilon \right) L\left( -\varepsilon \right) \cos \alpha
\left( 1+\sin ^{2}\alpha \right) \cos \alpha ^{\prime }\sin ^{2}\alpha
^{\prime }  \notag
\end{eqnarray}

The function $X_{\varepsilon }\left( \alpha ,\alpha ^{\prime }\right) $ is
determined from Eq. (\ref{0a48}) and we obtain :

\begin{eqnarray}
&&\left\vert X_{\varepsilon }\left( \alpha ,\alpha ^{\prime }\right)
\right\vert ^{2}=\sin ^{2}\beta -2\sin \beta \Big[L\left( \varepsilon
\right) \sqrt{1+\sin ^{2}\alpha }  \notag \\
&&\times \sin \left( \beta +\Phi \left( \alpha \right) \right) -L\left(
-\varepsilon \right) \cos \alpha ^{\prime }\cos \left( \beta +\Phi \left(
\alpha ^{\prime }\right) \right) \Big]  \notag \\
&&+2L\left( \varepsilon \right) L\left( -\varepsilon \right) \sqrt{1+\sin
^{2}\alpha }\cos \alpha ^{\prime }  \notag \\
&&\times \Big[\sin \left( 2\beta +\Phi \left( \alpha \right) +\Phi \left(
\alpha ^{\prime }\right) \right) -2\cos \phi \left( \alpha \right) \sin \Phi
\left( \alpha ^{\prime }\right) \Big]  \notag \\
&&+L^{2}\left( \varepsilon \right) \left( 1+\sin ^{2}\alpha \right)
+L^{2}\left( -\varepsilon \right) \cos ^{2}\alpha ^{\prime }  \label{0a56}
\end{eqnarray}

\subsection{Conductance $G_{NN}$ of the interface between two normal metals}

The transmission amplitude $t_{12}\left( \varepsilon \right) $ determines
the zero temperature conductance $G_{NN}\left( \varepsilon \right) $ between
two normal metals entering Eq. (\ref{0a10}). At a fixed $E_{F}$ the
conductance $G_{NN}\left( \varepsilon \right) $ is given by the following
formula

\begin{equation}
G_{NN}\left( \varepsilon \right) =G_{0}K_{0}\left( \varepsilon \right)
\int_{-\pi /2}^{\pi /2}\left\vert t_{12}\left( \varepsilon \right)
\right\vert ^{2}\cos \alpha d\alpha ,  \label{0a57}
\end{equation}

The final form of the condactance $G_{NN}\left( \varepsilon \right) $
between the normal metals can be written as

\begin{eqnarray}
G_{NN}\left( \varepsilon \right) &=&G_{0}K_{0}\left( \varepsilon \right)
\int_{-\pi /2}^{\pi /2}4L\left( \varepsilon \right) \cos ^{2}\alpha \left(
1+\sin ^{2}\alpha \right) d\alpha  \notag \\
&=&\frac{5\pi }{2}G_{0}K_{0}\left( \varepsilon \right) L\left( \varepsilon
\right)  \label{0a58}
\end{eqnarray}%
The formulas presented in this Section describe the conductances for all
parameters of interest and one can explicitly compare the corresponding
numerical curves with experimental results. All the details of the
derivation of the results of the present Section can be found in Appendix.

\begin{figure}[tbp]
\centering
\includegraphics[width=0.5\textwidth]{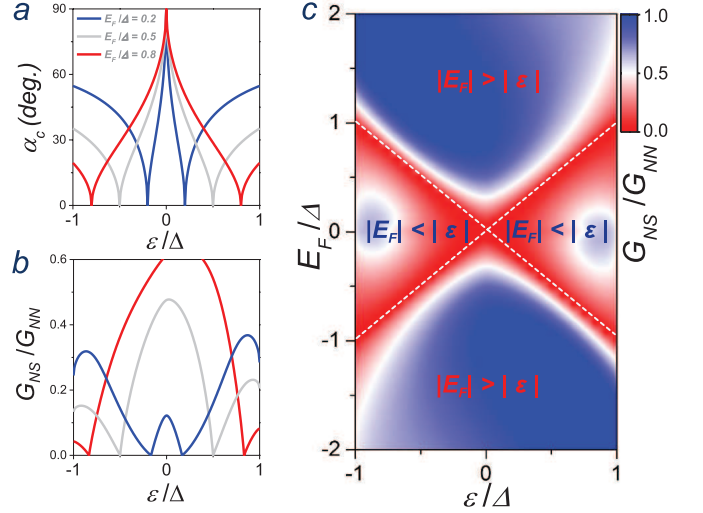}
\caption[Conductance for varying interfacial potential $U_{0}$]{(a) Angle of
total internal reflection for ARs $\protect\alpha_{c}$ for various fixed $%
E_{F}$. For $\protect\varepsilon = 0$, $\protect\alpha _{c}=\protect\pi/2$,
and ARs are possible for all angles of incidence. For $\protect\varepsilon =
E_{F}$, AR are prohibited for all angles and $\protect\alpha _{c}=0$. (b)
Conductance $G_{NS}/G_{NN}(10K)$ as a function as a function of $\protect%
\varepsilon$ for the same $E_{F}$ as in (a). The conductance is pinched off
at same positions at which $\protect\alpha _{c}=0$. (c) 2D colormap of $%
G_{NS}/G_{NN}(10K)$ for a function of both $\protect\varepsilon$ and $E_{F}$%
. The conductance dips scale along the diagonal lines defined by the
condition $|\protect\varepsilon| = |E_{F}|$ and define a phase diagram that
separates the map into regions of retro- and specular-ARs. Here we used the
parameter $U_{0}=5$meV.}
\end{figure}

\section{Numerical study and discussion}

In the AR process the incident electron has a total energy of $%
E_{F}+\varepsilon $ and condenses into a Cooper pair with a total energy of $%
2E_{F}$ after transmission into the SC. Here energy conservation dictates
that the reflected hole has a slightly lower total energy of $%
E_{F}-\varepsilon $. This energy loss of $2\varepsilon $ can have dramatic
consequences for an ultra-low $E_{F}$, where for the condition $|\varepsilon
|>E_{F}$, the incident conduction hole electron can be reflected as a hole
in the valence band (Fig. 1 (a) and (b).

Due to momentum conservation parallel to the interface $q$, the reduced
energy of the hole directly translates into an altered angle of reflection $%
\alpha ^{\prime }$ as compared to the angle of incidence $\alpha $. As was
shown in the previous section, these angles are related to each other by the
simple relation (\ref{0a43}) from which one can conclude that ARs are only
possible for angles $\alpha <\alpha _{c}$, where $\alpha _{c}$, Eq. (\ref%
{0a44}), plays the role of the angle of total internal reflection for ARs
and is plotted in Fig. 2 (a) for various fixed $E_{F}$. As can be seen from
the plots, for $\varepsilon =0$, $\alpha _{c}=\pi /2$, and ARs are possible
for all angles of incidence. However when $\varepsilon =E_{F}$, $\alpha
_{c}=0$, and AR are prohibited for all angles.

The explicit analytical formula for $G_{NS}$, Eq. (\ref{0a46}), and
subsequent formulas for the functions entering this equation allows us to
numerically calculate $G_{NS}(\varepsilon ,E_{F})$ for the full phase space.
Here we explicitly calculate the normalized conductance, $G_{NS}(\varepsilon
,E_{F})/G_{NN}(\varepsilon ,E_{F},T)$, since this allows to better highlight
the conductance features that arise solely from the SC proximity effect. As
both $G_{NS}$ and $G_{NN}$ in a realistic sample are generally subjected to
effect of energy fluctuations and disorder, one can eliminate these
undesired and hard to quantify effects by simply dividing these out. Since
measurements of the SC state are typically performed at $T\ll T_{c}$ \cite%
{dima}, one can use $G_{NS}(\varepsilon ,E_{F})$ at $T=0$, with a reasonable
accuracy. However, we specifically derive a temperature broadened form of $%
G_{NN}(\varepsilon ,E_{F},T)$ since in an experiment one can only obtain
this quantity at elevated temperatures $T>T_{c}$

\begin{equation}
G_{NN}(\varepsilon ,E_{F},T)=\frac{5\pi }{4}\int_{-\infty }^{\infty }\frac{%
K_{0}(\varepsilon )L(\varepsilon )dE_{F}}{4T\cosh ^{2}(\frac{%
E_{F}-\varepsilon }{2T})}.  \label{m4}
\end{equation}

In Fig. 2 (b) we plot out $G_{NS}/G_{NN}(10K)$ (here we choose $T=10$K for $%
G_{NN}$ as was used in Ref. \cite{dima}) as a function of $\varepsilon $ for
various fixed $E_{F}$. Since $\alpha _{c}=0$ for $|\varepsilon |=|E_{F}|$,
it can be seen from Eq. (\ref{0a46}) that the conductance vanishes, $G_{NS}=0$%
, at these points. As discussed earlier, this condition coincides exactly
with the condition that separates intra-band retro- from inter-band specular
reflections. The depleted interfacial resistance for this energy condition
marks therefore the cross-over between the two different regimes, and so
provides a strong experimental observable. In Fig. 2 (c) we plot a 2D
colormap of $G_{NS}/G_{NN}$ for a function of both $\varepsilon $ and $E_{F}$%
. The conductance dips scale along the diagonal lines defined by the
condition $|\varepsilon |=|E_{F}|$ and define a striking cross-like shape.
One can use this map as a phase diagram to separate the region of retro- and
specular-ARs.

The integrand in Eqs. (\ref{0a7}, \ref{0a46}) contains only one unknown
parameter, the potential step at the BLG/SC interface $U_{0}$ (in Fig. 2 we
used a $U_{0}=5$meV). In principle, this value can be found in literature
for various materials or can be extracted by fitting the experimental curves
with the presented theory. In general, it is important to emphasize the role
of $U_{0}$ for the purpose of experiments, as it can strongly affect the
outcome. In Fig. 3 we study the normalized conductance $G_{NS}/G_{NN}$ for
various values of $U_{0}$. One can see that the inner gap conductance is
strongly reduced for higher values of $U_{0}$. This results in the strong
suppression of the sharpness and contrast of the characteristic cross-over
point from retro- to specular ARs, $E_{F}=\varepsilon $. This suppression
can be explained by the previously discussed decaying transmission amplitude
$t_{12}(\varepsilon )$ which scales proportionally to $U_{0}^{-1/2}$ and is
therefore very small for large $U_{0}$. For experimental studies of specular
AR in BLG/SC junctions it is therefore important to engineer an interface
with a rather small potential step $U_{0}$.

\begin{figure}[tbp]
\centering
\includegraphics[width=0.5\textwidth]{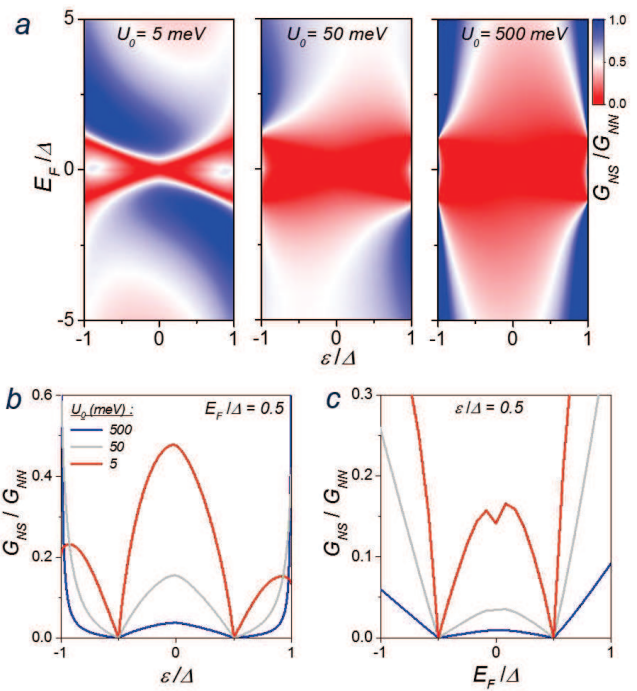}
\caption[Conductance for varying interfacial potential $U_{0}$]{Normalized
conductance $G_{NS}/G_{NN}(10K)$ for 3 different values of the potential
step at the N/SC interface $U_{0}$. As can be seen from the various graphs,
the inner gap conductance is strongly reduced for higher values of $U_{0}$.
This results in the strong suppression of the sharpness and contrast of the
characteristic cross-over point from retro- to specular ARs, $E_{F}=\protect%
\varepsilon$. (a) 2D colorplots as a function of $\protect%
\varepsilon$ and $E_{F}$. (b) Excitation energy dependence $\protect\varepsilon$ for a
fixed $|E_{F}| < \Delta$. (c) Fermi energy dependence $E_{F}$ for a fixed $|%
\protect\varepsilon| < \Delta$.}
\end{figure}

\section{Conclusions}

In this manuscript we have rigorously worked out analytical expressions for
the differential conductance across a BLG/SC interface for the ultimate
limits of large and small Fermi energies $E_{F}$. We have found, that while
for $E_{F} \gg \Delta$ the AR process is described by intra-band
retro-reflections, in the limit of $E_{F} \ll \Delta$ they are described by
inter-band specular-reflections. From numerical calculations we find that
the cross-over between the two processes has a clear experimental signature
that manifests itself in a strongly suppressed interfacial conductance when
the excitation energy $|\varepsilon| = |E_{F}| < \Delta$.

\section{Acknowledgements}

The authors thank C. W. J. Beenakker, P. Kim and J. D. Pillet for helpful
discussions. K.B. Efetov gratefully acknowledges the financial support of
the Ministry of Education and Science of the Russian Federation in the
framework of Increase Competitiveness Program of NUST \textquotedblleft
MISiS\textquotedblright\ (Nr. K2-2014-015) and Priority Program 1459
\textquotedblleft Graphene\textquotedblright\ of \textit{Deutsche
Forschungsgemeinschaft.}

\section{Appendix : Eigenenergies, wave functions, transmission and
reflection amplitudes in the absence of a SC gap}

Equations (\ref{0a7}-\ref{0a10}) allow one to reduce the calculation of the
conductance of the N/SC junction to study of the transmission and reflection
amplitudes between two normal metals. This can be done putting $\Delta=0$
in Eqs. (\ref{0a1}-\ref{0a2}) and writing the obtained equation in the form :

\begin{equation}
\left( \mathcal{H}_{0}(\mathbf{\hat{K}})+U(x)\right) u=Eu,  \label{k1}
\end{equation}%
with $U(x)$ defined in Eq. (\ref{0a3}). One should now solve Eqs. (\ref{k1})
separately at $x>0$ and $x<0$ and match the solutions at $x=0$.

\subsection{Eigenenergies}

First, following Refs. \cite{wallace,falko,guinea}, we write the
eigenenergies $E(\mathbf{K})$ of the Hamiltonian $\mathcal{H}_{0}(\mathbf{K})
$, Eq. (\ref{0a2}), for an arbitrary relation between $t_{\perp }$ and the
characteristic energies inside one layer in the form :

\begin{eqnarray}
E_{1}\left( \mathbf{K}\right)  &=&\hslash v\left( -t_{\perp }/2-\mathcal{E}%
\left( \mathbf{K}\right) \right) ,  \label{k2} \\
E_{2}\left( \mathbf{K}\right)  &=&\hslash v\left( -t_{\perp }/2+\mathcal{E}%
\left( \mathbf{K}\right) \right) ,  \notag \\
E_{3}\left( \mathbf{K}\right)  &=&\hslash v\left( t_{\perp }/2+\mathcal{E}%
\left( \mathbf{K}\right) \right) ,  \notag \\
E_{4}\left( \mathbf{K}\right)  &=&\hslash v\left( t_{\perp }/2-\mathcal{E}%
\left( \mathbf{K}\right) \right) ,  \notag
\end{eqnarray}%
where

\begin{equation}
\mathcal{E}\left( \mathbf{K}\right) =\sqrt{\left( t_{\perp }/2\right)
^{2}+K^{2}}  \label{k3}
\end{equation}

Provided the interlayer coupling $t_{\perp}$ exceeds all the other energies,
the bands of the Hamiltonian (\ref{0a2}) with the spectra $E_{1}\left(
\mathbf{k}\right) $ and $E_{3}\left( \mathbf{k}\right) $ are far away from
the Fermi energy and their contribution into physical quantities can be
neglected. The eigenenergies of the first two low energy bands take in the
limit of small $\left\vert \mathbf{k}\right\vert \ll t_{\perp }$ the
following form:

\begin{eqnarray}
E_{2}\left( \mathbf{k}\right) &=&\hslash v\left( -t_{\perp }/2+\mathcal{E}%
\left( \mathbf{k}\right) \right) \approx \frac{\hslash vK^{2}}{t_{\perp }},
\label{k4} \\
E_{4}\left( \mathbf{k}\right) &=&\hslash v\left( t_{\perp }/2-\mathcal{E}%
\left( \mathbf{k}\right) \right) \approx -\frac{\hslash vK^{2}}{t_{\perp }}.
\notag
\end{eqnarray}

In Eq. (\ref{k4}), $E_{2}\left( \mathbf{K}\right) $ describes the conduction
band and $E_{4}\left( \mathbf{K}\right) $ describes the valence band. Using
the inequality (\ref{0a5}) we consider only these low-lying bands and no
higher energy bands.

\subsection{Wave functions}

The Hamiltonian $\mathcal{H}_{0}\left( \mathbf{k}\right) $, Eq. (\ref{0a2}),
can be diagonalized as was done in Ref. \cite{guinea}, and we can easily
obtain $4$-component vectors $u$ that satisfy the equation

\begin{equation}
\left( \mathcal{H}\left( \mathbf{k}\right) -E_{F}\right) u=\varepsilon u.
\label{k5a}
\end{equation}

Here we consider only the case $E_{F}>0$. The solutions depend on the sign
of $\varepsilon +E_{F}$ and we write them separately for $\varepsilon
+E_{F}>0$ and $\varepsilon +E_{F}<0$.

For the low-lying eigenvalues we obtain

\begin{eqnarray}
\varepsilon _{2} &=&-E_{F}+\frac{\hslash v}{t_{\perp }}K^{2},\quad
\label{0a21} \\
\varepsilon _{4} &=&-E_{F}-\frac{\hslash v}{t_{\perp }}K^{2}  \label{0a19}
\end{eqnarray}

In order to calculate the wave functions one should choose an eigenvalue $%
\varepsilon $ and determine $K$ as a function of this $\varepsilon $. It is
clear that constructing plain waves in the region of the normal metal $N_{1}$
and $E_{F}>0$ one should take the solution of Eq. (\ref{0a21}) for $K$ as a
function of $\varepsilon $ at $\varepsilon > -E_{F}$ and of Eq. (\ref{0a19})
at $\varepsilon < -E_{F}$. In addition, one has a solution for $K$ of (\ref%
{0a19}) at $\varepsilon > -E_{F}$ and of (\ref{0a21}) at $\varepsilon <
-E_{F}$. The latter solutions are either exponentially growing or decaying
as functions of $x$. Nevertheless, they should also be taken into account
when matching functions at the interface because we have the deep potential $%
-U_{0}$ at $x<0$ and the exponential growth can change to the plain wave
behavior there.

Here we calculate the wave functions for the region $x>0$. These can be
however also used in the region $x<0$ after shifting the energy $%
E_{F}\rightarrow E_{F}+U_{0}.$We will later on denote all quantities where
this shift has been done by adding the subscript $U_{0}$, thus obtaining $%
u_{1U_{0}}^{R,L},$ $u_{2U_{0}}^{R,L},$ etc.

\subsubsection{Plain wave solutions at $\protect\varepsilon +E_{F}>0$}

In this region of the energies we have left and right moving electrons of
the conduction band and use Eq. (\ref{0a21}). The solution $u_{1}^{R}$ for
right moving particles in this region takes the form :

\begin{equation}
u_{1}^{R}=\frac{e^{ikx+iqy}}{2\sqrt{K_{0}/t_{\perp }\cos \alpha }}\left(
\begin{array}{c}
K_{0}/t_{\perp } \\
e^{-i\alpha } \\
K_{0}/t_{\perp } \\
e^{i\alpha }%
\end{array}%
\right) ,\quad   \label{0a13}
\end{equation}%
while the solution for the left moving particles $u_{1}^{L}$ reads (we use a
compact notation $K_{0}=K_{0}\left( \varepsilon \right) $, Eq. (\ref{e2}))

\begin{equation}
u_{1}^{L}=\frac{e^{-ikx+iqy}}{2\sqrt{K_{0}/t_{\perp }\cos \alpha }}\left(
\begin{array}{c}
-K_{0}/t_{\perp } \\
e^{i\alpha } \\
-K_{0}/t_{\perp } \\
e^{-i\alpha }%
\end{array}%
\right) .  \label{0a14}
\end{equation}

The wave functions $u_{1}^{R}$ and $u_{1}^{L}$ correspond to the eigenenergy
$E_{2}\left( \mathbf{k}\right) $, Eq. (\ref{k4}), and belong to the
conduction band. They are normalized assuming the current $1$ along the $x$%
-axis for the right moving particles and $-1$ for left moving ones. This
fact can easily be checked using the matrix form of the current operator

\begin{equation}
\mathbf{j=}e\mathbf{\sigma }  \label{k6h}
\end{equation}%
having the $x$-component

\begin{equation}
j_{x}=e\sigma _{x},  \label{k6a}
\end{equation}%
where $\mathbf{\sigma }$ is the vector of Pauli matrices in the sublattice
space of graphene.

Having fixed $E_{F}>0$ we have to express $k$ and $q$ in terms of $K_{0}$
and $\alpha $. Since we introduced the angle $\alpha $ as

\begin{equation}
k-iq=K_{0}e^{-i\alpha },  \label{0a17}
\end{equation}%
we obtain for $\varepsilon +E_{F}>0$ the following relations for the
variables $\alpha $ and $k$

\begin{equation}
k=K_{0}\cos \alpha ,\quad q=K_{0}\sin \alpha \quad   \label{0a18}
\end{equation}%
Here the angle $\alpha $ varies in the interval $-\pi /2<\alpha <\pi /2$.

\subsubsection{Plain wave solutions at $\protect\varepsilon +E_{F}<0$}

In this region of energies we have plain waves corresponding to right and
left moving holes from the valence band and use Eq. (\ref{0a19}). For the
right moving holes we obtain the normalized wave functions :

\begin{equation}
u_{2}^{R}=\frac{e^{ik^{\prime }x+iqy}}{2\sqrt{K_{0}/t_{\perp }\cos \alpha
^{\prime }}}\left(
\begin{array}{c}
K_{0}/t_{\perp } \\
e^{-i\alpha ^{\prime }} \\
-K_{0}/t_{\perp } \\
-e^{i\alpha ^{\prime }}%
\end{array}%
\right) ,\quad   \label{0a22}
\end{equation}%
where

\begin{equation}
k^{\prime }=-K_{0}\cos \alpha ^{\prime },\quad q=-K_{0}\sin \alpha ^{\prime
}.  \label{0a24}
\end{equation}

In Eqs. (\ref{0a22}, \ref{0a24}), the angle $\alpha ^{\prime }$ varies in
the interval $-\pi /2<\alpha ^{\prime }<\pi /2.$

The opposite signs in Eq. (\ref{0a24}), as compared to Eq. (\ref{0a18}), are
due to the fact that we now consider holes from the valence band instead of
electrons from the conduction band.

The solution $u_{2}^{L}$ for the left moving particles takes the form

\begin{equation}
u_{2}^{L}=\frac{e^{-ik^{\prime }x+iqy}}{2\sqrt{K_{0}/t_{\perp }\cos \alpha
^{\prime }}}\left(
\begin{array}{c}
-K_{0}/t_{\perp } \\
e^{i\alpha ^{\prime }} \\
K_{0}/t_{\perp } \\
-e^{-i\alpha ^{\prime }}%
\end{array}%
\right) .  \label{0a26}
\end{equation}

The current of the right moving holes (functions $u_{2}^{R}$) equals $1,$
while the current for the left moving holes (function $u_{2}^{L}$) equals $%
-1 $.

\subsubsection{Decaying and growing solutions at $\protect\varepsilon %
+E_{F}<0$}

In addition to the plain waves, Eqs. (\ref{0a13}, \ref{0a14}), there are two
other solutions $u_{1}^{<}$ and $u_{1}^{>}$ corresponding to the eigenvalue $%
\varepsilon _{2}$, Eq. (\ref{0a21}) from the conduction band :

\begin{equation}
u_{1}^{<}=e^{\kappa x+iqy}\left(
\begin{array}{c}
-iK_{0}/t_{\perp } \\
e^{\gamma } \\
-iK_{0}/t_{\perp } \\
e^{-\gamma }%
\end{array}%
\right)  \label{0a15}
\end{equation}

and

\begin{equation}
u_{1}^{>}=e^{-\kappa x+iqy}\left(
\begin{array}{c}
iK_{0}/t_{\perp } \\
e^{-\gamma } \\
iK_{0}/t_{\perp } \\
e^{\gamma }%
\end{array}%
\right)  \label{0a16}
\end{equation}

The normalization in Eq. (\ref{0a15}, \ref{0a16}) does not play any role and
we just set it equal to unity. The parameters $\kappa $ and $q$ can be
written in the form :

\begin{equation}
\kappa =K_{0}\cosh \gamma ,\quad q=K_{0}\sinh \gamma  \label{0a27}
\end{equation}

\subsubsection{Decaying and growing solutions at $\protect\varepsilon %
+E_{F}>0$}

In this region of parameters there are growing and decaying wave functions
that correspond to $\varepsilon _{4}$ from Eq. (\ref{0a19}) and belong to
the valence band. We write the growing $u_{2}^{<}$ and decaying $u_{2}^{>}$
functions as

\begin{equation}
u_{2}^{<}=e^{\kappa x+iqy}\left(
\begin{array}{c}
iK_{0}/t_{\perp } \\
e^{\gamma ^{\prime }} \\
-iK_{0}/t_{\perp } \\
-e^{-\gamma ^{\prime }}%
\end{array}%
\right)   \label{0a28}
\end{equation}%
and

\begin{equation}
u_{2}^{>}=e^{-\kappa x+iqy}\left(
\begin{array}{c}
-iK_{0}/t_{\perp } \\
e^{-\gamma ^{\prime }} \\
iK_{0}/t_{\perp } \\
-e^{\gamma ^{\prime }}%
\end{array}%
\right)  \label{0a29}
\end{equation}

The parameters $\kappa $ and $q$ can be written in the form

\begin{equation}
\kappa =K_{0}\cosh \gamma ^{\prime },\quad q=K_{0}\sinh \gamma ^{\prime }
\label{0a30}
\end{equation}

\subsubsection{Relations between the variables $\protect\alpha ,\protect%
\alpha^{\prime }$ and $\protect\gamma^{\prime },\protect\gamma$}

The variables $\alpha ,\alpha ^{\prime },\gamma $ and $\gamma ^{\prime }$
are not independent because the component $q$ parallel to the interface is
everywhere the same. Comparing Eqs. (\ref{0a18}) and (\ref{0a30}) we come to
the relation :

\begin{equation}
\sin \alpha =\sinh \gamma ^{\prime },  \label{0a30a}
\end{equation}%
while the comparison of Eqs. (\ref{0a24}) and (\ref{0a27}) leads to

\begin{equation}
\sin \alpha ^{\prime }=-\sinh \gamma  \label{0a30b}
\end{equation}

\subsection{Transmission $t_{21}\left( \protect\varepsilon \right) $, $%
t_{12}\left( \protect\varepsilon \right) $ and reflection $r_{21}\left(
\protect\varepsilon \right) $ amplitudes}

Now we calculate the transmission $t_{21}\left( \varepsilon \right) $ and
reflection $r_{21}\left( \varepsilon \right) $ amplitudes that match the
wave functions written on the left and on the right from the interface.
Again, we consider the regions $\varepsilon +E_{F}>0$ and $\varepsilon
+E_{F}<0$ separately. Then, the transmission amplitude $t_{12}\left(
\varepsilon \right) $ can easily be obtained from $t_{21}\left( \varepsilon
\right) $ through the well know relation

\begin{equation}
t_{12}\left( \varepsilon \right) =t_{21}\left( \varepsilon \right)
e^{i\delta \left( \varepsilon \right) },  \label{0a42}
\end{equation}%
where $\delta \left( \varepsilon \right) $ is a phase whose explicit value
is not necessary for the calculation of the conductances using Eqs. (\ref%
{0a7}-\ref{0a10}).

\subsubsection{Region $E_{F}>0$, $\protect\varepsilon +E_{F}>0$}

The scattering process at $x<0$ includes a plane wave that is incident from
the left, $u_{1U_{0}}^{R},$ and another one, $u_{1U_{0}}^{L},$ reflected
from the interface. In addition, in the region $x<0$ one has a wave that is
growing with $x$ (decaying from the interface) with the symmetry of $%
u_{1}^{<}$ from Eq. (\ref{0a15}).

After scattering off the interface one obtains an outgoing wave for $x>0$
with the structure $u_{1}^{R},$ Eq. (\ref{0a13}), and a decaying wave with
the structure $u_{2}^{>}$, Eq. (\ref{0a29}). We describe the scattering
process for $\varepsilon +E_{F}>0$ that matches these functions at the
interface :

\begin{equation}
u_{1U_{0}}^{R}+r_{22}\left( \varepsilon \right)
u_{1U_{0}}^{L}+Bu_{2U_{0}}^{<}=t_{21}\left( \varepsilon \right)
u_{1}^{R}+Cu_{2}^{>},  \label{0a31}
\end{equation}%
where $B$ and $C$ are coefficients have to be found by solving Eq. (\ref%
{0a31}). Actually, Eq. (\ref{0a31}) is a system of $4$ linear equations and
one has to find four unknown quantities $r_{22}\left( \varepsilon \right) ,$
$t_{21}\left( \varepsilon \right) ,$ $B\left( \varepsilon \right) $ and $%
C\left( \varepsilon \right) $.

Writing Eqs. (\ref{0a31}) explicitly leads to some quite cumbersome
expressions. Fortunately, these equations are simpler in the limit $U_{0}\gg
\varepsilon ,E_{F}$, which is explicitly considered here. The amplitudes $%
t_{21}\left( \varepsilon \right) $ and $r_{22}\left( \varepsilon \right) $
can now be found rather easily. Using Eq. (\ref{e1}) we write

\begin{equation}
\frac{\sin \alpha \left( U_{0}\right) }{\sin \alpha }\simeq \sqrt{\frac{%
\varepsilon +E_{F}}{U_{0}}}=L\left( \varepsilon \right) \ll 1  \label{k20e}
\end{equation}%
to obtain in the linear approximation in $\alpha $

\begin{equation}
\cos \alpha \left( U_{0}\right) =\sqrt{1-L^{2}\left( \varepsilon \right)
\sin ^{2}\alpha }\simeq 1.  \label{k20f}
\end{equation}

We can hence approximate

\begin{equation}
e^{i\alpha \left( U_{0}\right) }\simeq 1+iL\left( \varepsilon \right) \sin
\alpha ,\quad e^{\gamma \left( U_{0}\right) }=1+L\left( \varepsilon \right)
\sinh \gamma .  \label{k20g}
\end{equation}%
and simplify Eqs. (\ref{0a31}) with the help of the relations (\ref{k20e}-%
\ref{k20g}). We solve these equations and arrive at the following expression
for the transmission coefficients $t_{21}\left( \varepsilon \right) $ and $%
r_{22}\left( \varepsilon \right) $ :

\begin{equation}
t_{21}\left( \varepsilon \right) =\frac{2\sqrt{L\left( \varepsilon \right)
\cos \alpha }\cosh \gamma ^{\prime }}{\cos \alpha \cosh \gamma ^{\prime
}+i\sin \alpha \sinh \gamma ^{\prime }},  \label{m13}
\end{equation}%
and

\begin{equation}
r_{22}\left( \varepsilon \right) =1-\frac{2L\left( \varepsilon \right) \cosh
\gamma ^{\prime }}{\cos \alpha \cosh \gamma ^{\prime }+i\sin \alpha \sinh
\gamma ^{\prime }}  \label{m14}
\end{equation}

Using Eq. (\ref{k20e}) we rewrite Eqs. (\ref{m13}-\ref{m14}) in a more
compact form

\begin{equation}
t_{21}\left( \varepsilon \right) =2\sqrt{L\left( \varepsilon \right) \cos
\alpha \left( 1+\sin ^{2}\alpha \right) }\exp \left( -i\Phi \left( \alpha
\right) \right) ,  \label{0a33}
\end{equation}%
with $\Phi \left( \alpha \right) $ introduced in Eq. (\ref{0a45}).

Since the angle $\Phi \left( \alpha \right) $ varies in the interval

\begin{equation}
0<\Phi \left( \alpha \right) <\pi /2  \label{0a35}
\end{equation}%
we obtain for the reflection coefficient

\begin{equation}
r_{22}\left( \varepsilon \right) =1-2L\left( \varepsilon \right) \sqrt{%
\left( 1+\sin ^{2}\alpha \right) }\exp \left( -i\Phi \left( \alpha \right)
\right)  \label{0a36}
\end{equation}

The unitarity condition immediately follows from Eqs. (\ref{0a33}, \ref{0a36}%
) in the limit (\ref{k20e})

\begin{equation}
\left\vert t_{21}\left( \varepsilon \right) \right\vert ^{2}+\left\vert
r_{22}\left( \varepsilon \right) \right\vert ^{2}=1  \label{0a37}
\end{equation}

\subsubsection{Region $E_{F}>0$, $\protect\varepsilon +E_{F}<0$}

In the region $E_{F}>0,$ $\varepsilon +E_{F}<0$ matching the wave functions
at $x=0$ results in the equation :

\begin{equation}
u_{1U_{0}}^{R}+r_{22}\left( \varepsilon \right) u_{1U_{0}}^{L}+B\left(
\varepsilon \right) u_{2U_{0}}^{<}=t_{21}\left( \varepsilon \right)
u_{2}^{R}+C\left( \varepsilon \right) u_{1}^{>},  \label{0a38}
\end{equation}

Using the same approximation (\ref{k20e}-\ref{k20g}) and proceeding in the
same way as for $\varepsilon +E_{F}$ one comes with the help of Eq. (\ref%
{0a30b}) to the following result for the transmission amplitude

\begin{equation}
t_{21}\left( \varepsilon \right) =-2\sqrt{L\left( \varepsilon \right) \cos
\alpha ^{\prime }}\sin \alpha ^{\prime }\exp \left( i\Phi \left( \alpha
^{\prime }\right) \right)  \label{0a39}
\end{equation}

The reflection amplitude $r_{22}\left( \varepsilon \right) $ can be then
written as

\begin{equation}
r_{22}\left( \varepsilon \right) =1-2iL\left( \varepsilon \right) \cos
\alpha ^{\prime }\exp \left( i\Phi \left( \alpha ^{\prime }\right) \right) ,
\label{0a40}
\end{equation}%
with

\begin{equation}
\Phi \left( \alpha ^{\prime }\right) =-\arcsin \left( \sin ^{2}\alpha
^{\prime }\right)  \label{0a41}
\end{equation}

Again, the unitarity condition, Eq. (\ref{0a37}), is fulfilled in the limit
specified in Eq. (\ref{k20e}).


\begin{thebibliography}{99}
\bibitem{mention} We want to mention that the conductance across a BLG/SC interface was
already calculated in Ref. \onlinecite{ludwig}. However, the calculations in
that publication were done assuming that the potential $U_{0}$ is highest energy in the system and explicit
results were written only in some limiting case. In contrast, we use a more
relevant assumption that the potential $U_{0} $ is smaller
than the interlayer coupling energy and derive general formulas that allow
us to make the comparison with the experiment \cite{dima} in the entire
region of the parameters $\varepsilon $ and $E_{F}$
\end{thebibliography}
\end{document}